\documentclass[superscriptaddress,secnumarabic,twocolumn,
 amssymb,amsmath,nobibnotes,aps,prd,showkeys,showpacs]{revtex4}
\usepackage{graphicx,epsfig,amssymb,amsmath,amstext}
\usepackage{bm}
\usepackage[latin1]{inputenc}
\usepackage[T1]{fontenc} 
\usepackage[english]{babel}
\usepackage{lmodern}

\usepackage{amsmath}
\usepackage{amssymb}
\usepackage{booktabs}
\usepackage{color}
\usepackage{nicefrac}
\usepackage{dsfont}
\newcommand{\R}{\mathbb{R}} 
 
\newcommand{\C}{\mathbb{C}}

\newcommand{\ii}{\mathrm i}

\newcommand{\dt}[1]{\frac{\mathrm d #1}{\mathrm d t}} % for derivatives
\usepackage{hyperref}
\usepackage{verbdef}
\usepackage[vlined,ruled]{algorithm2e}
 \newcommand\BlockIf[1]{#1}
 \newcommand\BlockElseIf[1]{#1}
 \newcommand\BlockElse[1]{#1}
\makeatletter
\renewcommand{\uIf}[2]{\If@ifthen{#1}\If@noend{\BlockIf{#2}}}
\renewcommand{\uElseIf}[2]{\ElseIf@elseif{#1}\If@noend{\BlockElseIf{#2}}}
\renewcommand{\Else}[1]{\Else@else\If@endif{\BlockElse{#1}}}
\makeatother

\usepackage{xcolor}
\definecolor{color1}{rgb}{.8,.8,.8}
\definecolor{color2}{rgb}{.65,.65,.65}
\definecolor{color3}{rgb}{.35,.35,.35}
\definecolor{color4}{rgb}{0,0,0}

\begin{document}
\title{Splitting integrators for the BCS equations of superconductivity}
\author{Jonathan Seyrich}
\email{seyrich@na.uni-tuebingen.de}
\affiliation{Mathematisches Institut, Universit\"{a}t T\"{u}bingen,
Auf der Morgenstelle, 72076 T\"{u}bingen, Germany}
\begin{abstract}
The BCS equations are the centerpiece of the microscopic description of superconductivity.
Their space discretization yields a system of coupled ordinary differential equations. In this work, we come up with 
fast time evolution schemes based on a splitting approach. One of the schemes only requires basic operations. For the 
physically important case of the BCS equations for a contact interaction potential,
the computational cost of the schemes grows
only linearly with the dimension of the space discretization. Their accuracy is demonstrated
in extensive numerical experiments. These experiments also show that the physical energy of the system is preserved
up to very small errors.
\end{abstract}

\maketitle
%%%%%%%%%%%%%%%%%%%%%%%%%%%%%%%%%%%%%%%%%%%%%%%%%%%%%%%%%%%%%%%%%%%%%%%%%%%%%%%%%%%%%%
%%%%%%%%%%%%%%%%%%%%%%%%%%%%%%%%%%%%%%%%%%%%%%%%%%%%%%%%%%%%%%%%%%%%%%%%%%%%%%%%%%%%%%
\section{Introduction}\label{IntSec:Intro}

In this work, we consider the time-dependent BCS equations, often also referred to as Bogolubov--de-Gennes (BdG) equations.
These equations, named after \textit{Bardeen, Cooper and Schrieffer}, see~\cite{bardeen1957theory}, are 
are the basis of the microscopic description
of superconductivity in metals. They are coupled partial differential equations which describe 
the evolution of the particle density $\gamma$ and the Cooper pair density $\alpha$ of a fermionic system (see also~\cite{bach1994generalized}
for a detailed introduction). Although the BCS\slash BdG equations
are a fundamental part of condensed matter physics, their numerical treatment has not been paid attention to so far. Thus,
in this work, we present two reliable and efficient integration algorithms for these equations based on a splitting approach.

The evolution equations for the Cooper pair density
resembles the linear Schr\"odinger equation for quantum dynamical systems. 
One important aspect of these systems is that, after a space discretization, the right hand side of the resulting ordinary differential equations has a very large
Lipschitz constant caused by the Laplacian in the kinetic part. As a consequence, 
standard explicit integration schemes, such as the ones presented in~\cite{1992F&PT..8..NR}, although very popular in computational physics, 
are of no use in quantum mechanical applications. 
Therefore, the treatment of such quantum
dynamical systems has been of huge interest in the numerical analysis community for many decades, 
see,~e.g.,~\cite[Chapter~II.~1]{lubich2008quantum}.
Various evolution schemes for the linear
Schr\"odinger equation in varying settings have been proposed over the years, 
see, e.g.,~\cite{gray1996symplectic,blanes2006symplectic,tal1984accurate,park1986unitary,hochbruck1997krylov,hochbruck2003magnus}. 
Nonlinear Schr\"odinger equations such as the Gross--Pitaevskii equation and 
equations arising from the Hartree and Hartree--Fock approximation of the quantum state have also been devoted attention to, see,
e.g.,~\cite{tang1996symplectic,bao2003numerical,gauckler2010splitting}~and~\cite{lubich2004variational,lubich2005variational}, respectively. 

All theses methods have in common that the partial differential equations are first discretized in space. This means that 
the system is restricted to a suitable subspace spanned by a finite number of basis functions. Here, we do this with the help of
a \textit{Fourier collocation} method which is the straightforward approach for the problem we look at, 
see, e.g.,~\cite[Chapter~III.~1]{lubich2008quantum}. This yields a system of coupled ordinary differential equations, on the solution of
which we focus in the present work.

What, in many applications, turned out to be the most promising tool for the solution of the space discretized system
was the splitting of the equations under consideration into some subproblems, each of which can be solved
more easily than the system of equations as a whole. This idea was first employed
for advection equations in~\cite{strang1968construction}~and~\cite{marchuk1968some}. In the realm of quantum dynamics, it was applied for the first time 
in~\cite{feit1982solution} where
the linear Hamiltonian was split into a kinetic and a potential part.
The respective solutions were then concatenated in a suitable way
in order to obtain a reliable integration method.
Here, we use this ansatz to introduce two schemes for the evolution of the space discretized BCS equations.
The coupling terms depend on the convolution of the particle density with the Cooper pair density. We use
the \textit{fast Fourier transform (FFT)} to swiftly compute these terms.
As a consequence, the CPU effort per time step of our schemes grows only 
mildly with the number of basis functions. This is very important since in most physical
applications the BCS system requires a discretization space of very high dimension. 

For the first scheme, we exploit 
that the eigenvalues of the density operator, which are functions of the particle density and the Cooper pair density, 
are conserved along exact solutions to the BCS equations. Hence, we can express the particle density as a function of the Cooper pair density.
We end up with a decoupled nonlinear system for the evolution of the Cooper pair density $\alpha$. The thus obtained equations
are split into a linear part, which can be solved exactly, and into a nonlinear part, the flow of which can be approximated by some standard numerical scheme. In the 
rest of this work, we will refer to the resulting integrator as \textit{BCSInt}. It is very accurate and preserves the physically interesting eigenvalues
of the density operator by construction.
The integrator has already
been employed in a numerical study of the physical behavior of the BCS equations in~\cite{hainzl2015comparing}.

For the second integrator, we do not decouple the system at all. Instead,
thanks to the system's particular structure, we can aptly split it into three subproblems for which the flows can be
calculated very efficiently. These calculations require only basic operations. Recombining the thus obtained flows in a suitable way results in a very accurate
and efficient scheme which conserves the system's constants of motion, such as the energy, up to very small errors. In the following, we will denote 
the new scheme by \textit{SplitBCS}. In the physically important case of a contact interaction, i.e., when the potential is given by a delta function,
the flows of the subproblems can all be calculated exactly with an effort linear in the number of basis functions.

We demonstrate our integrators' favorable behavior 
with the help
of numerical experiments and numerical comparisons to standard integration schemes. 
We mention that an error estimate similar to the one in~\cite{gauckler2011convergence} for splitting schemes applied to the 
Gross--Pitaevskii equations is expected to hold for SplitBCS. However, such an analysis is out of the scope of this work.

Our presentation 
is organized as follows: We start with a short introduction to the BCS equations in Section~\ref{IntSec:Deriv}.
Afterwards, we explain the Fourier collocation for the partial differential equations in Section~\ref{IntSec:SpaceDisc}.
We first introduce our splitting scheme BCSInt for the decoupled nonlinear system in Section~\ref{IntSec:Nonlinear}. 
Then, we present our fast integration scheme SplitBCS for the coupled system in 
Section~\ref{IntSec:FastInt}. This is followed by numerical tests in Section~\ref{IntSec:Exp}. Finally, we summarize our results in Section~\ref{IntSec:Con}.
\section{The BCS Equations}\label{IntSec:Deriv}
A superconducting translation invariant system in one spatial dimension is characterized by the particle density $\gamma:\mathbb R\times\mathbb R\mapsto\mathbb R$
which describes the probability at time $t$ of finding a particle at position $x$
and the Cooper pair density $\alpha:\mathbb R\times\mathbb R\mapsto\mathbb C$ which gives the probability at time $t$ of having a Cooper pair 
of electrons at distance $x$. For a given particle interaction $V$, the evolution of $\alpha$ and $\gamma$
is governed by the BCS equations, sometimes also called Bogolubov--De-Gennes equations,
\begin{align}
 \ii\dot\gamma(t,x)&=-2\int_{\mathbb R}V(y)\operatorname{Im}\left[\alpha(t,x-y)\overline{\alpha(t,y)}\right]\mathrm dy,\\
 \ii\dot\alpha(t,x)&=2\left(-\frac{\mathrm d^2}{\mathrm dx^2}-\mu+V(x)\right)\alpha(t,x)\\
 &-4\int_{\mathbb R}\gamma(t,x-y)V(y)\alpha(t,y)\mathrm dy,\nonumber
\end{align}
with $\mu$ denoting the chemical potential of the physical system and $\dot{}=\nicefrac{\partial }{\partial t}$.
Conventionally, the BCS equations are given in terms of the Fourier transforms, i.e., the momentum space representations
\begin{align}
&\hat\gamma(t,p)=\frac1{2\pi}\int_{\mathbb R}\gamma(t,x)e^{\ii px}\mathrm dx,\\
&\hat\alpha(t,p)=\frac1{2\pi}\int_{\mathbb R}\alpha(t,x)e^{\ii px}\mathrm dx.
\end{align}
In this basis, the equations can be written in the compact, self-consistent form
\begin{align}\label{eqn-Int-BCS-compact}
  \ii\dot\Gamma(t,p)=\left[H_{\Gamma(t,p)},\Gamma(t,p)\right], \quad p\in\mathbb R,
\end{align}
see,~e.g.,~\cite{FHSchS}. $\Gamma(t,p)$ is the $2\times2$-matrix
\begin{align}
  \Gamma(t,p)=\begin{pmatrix}\hat\gamma(t,p)&\hat\alpha(t,p)\\\overline{\hat\alpha(t,p)}&1-\hat\gamma(t,p)\end{pmatrix}
\end{align}
and $H_{\Gamma(t,p)}$ is the Hamiltonian
\begin{align}
  H_{\Gamma(t,p)}(p)&=\begin{pmatrix}p^2-\mu&2[\hat V\ast\hat\alpha](t,p)\\2[\hat V\ast\overline{\hat\alpha}](t,p)&\mu-p^2\end{pmatrix}.
\end{align}
Here, $\ast$ denotes the convolution of $\hat V$ with $\hat\alpha(t,p)$.
\subsection{Superconductivity}
It can be shown, see~e.g.~\cite{hainzl2008bcs}, that the free energy functional
\begin{align}\label{eqn-Int-free-energy}
 F_T(\Gamma(t))=&\int_{\R}(p^2-\mu)\hat\gamma(t,p)\mathrm dp+\int_{\R} |\alpha(t,x)|^2V(x)\mathrm dx\nonumber\\
 &+\int_{\R}\operatorname{Tr}_{\C^2}\left(\Gamma(p)\log\Gamma(p)\right)\mathrm dp
\end{align}
is conserved along solutions of the evolution equations~\eqref{eqn-Int-BCS-compact} for any given temperature of the system $T$. 
If, for a given temperature $T$, the minimizer $\Gamma$ of $\mathcal F_T$ has a non vanishing Cooper pair density $\alpha$,
then the system is said to be in a superconducting state.

\subsection{The discrete BCS equations}
In order to render the system computationally palpable, one restricts it to a domain $D=[0,L2\pi]$, $L\in\mathbb N$, and assumes
periodic boundary conditions. In most applications, $L$ is a large
integer as the extension of the system is considered to be huge compared to the microscopic scale which here is $\mathcal O(1)$. On the finite domain
$D$, the momenta consist of the discrete set $k\in\nicefrac1L\mathbb Z$. The momentum space representations of $\alpha$ and $\gamma$ are 
given by
\begin{align}
&\hat\gamma_k(t)=\frac1{L2\pi}\int_0^{L2\pi}\gamma(t,x)e^{\ii kx}\mathrm dx,\\
&\hat\alpha_k(t)=\frac1{L2\pi}\int_0^{L2\pi}\alpha(t,x)e^{\ii kx}\mathrm dx.
\end{align}
In terms of these representations, the BCS equations read
\begin{align}\label{eqn-Int-BCS-k}
  \ii\dot\Gamma_k(t)=\left[H_{\Gamma_k(t)},\Gamma_k(t)\right], \quad k\in\frac1L\mathbb Z,
\end{align}
where the convolution appearing in the Hamiltonian is now to be understood as
\begin{align}
 \left(\hat V\ast\hat\alpha\right)_k(t)=\sum_{j\in\mathbb Z}\hat V_{k-j}\hat\alpha_j(t).
\end{align}

The first step to a numerical solution is to introduce a finite basis. This process is called \textit{space discretization}.
\section{Space Discretization}\label{IntSec:SpaceDisc}
As the BCS equations are given in their momentum space representation anyway, it is most convenient to use the so-called \textit{Fourier collocation}.
This means that for a fixed number $K\in\mathbb N$, a $L2\pi$-periodic
function $f(x)=\sum_{j\in\mathbb Z}\hat f(j)e^{\ii \nicefrac kLx}$ is approximated by
\begin{align}\label{eqn-Int-def-colloc}
  f^K(x)=\sum_{k=-\frac K2}^{\frac K2-1}\hat f^K_ke^{\ii \frac kLx},
\end{align}
where the coefficients $\hat f^K_k$ are obtained by the discrete Fourier 
transform of the values $f_j=f\left(\nicefrac{L2\pi}K\cdot j\right)$, $j=-\nicefrac K2,...,\nicefrac K2-1$. 
From numerical analysis, cf.~\cite[Chapter~III.1]{lubich2008quantum}, it is known that for an $s$-times differentiable function $f$, the bound
\begin{align}
\Vert f(x)-f^K(x)\Vert\le CK^{-s}\|\frac{\mathrm d^sf}{\mathrm dx^s}\|
\end{align}
holds for some constant $C$ independent of the number of basis functions $K$.

Mathematically speaking, we work on the subspace spanned by the first $K$ eigenfunctions of the Laplacian on $[0,L2\pi]$.
The approximation of the particle density on this subspace is given by
\begin{align}\label{eqn-Int-gamma-colloc}
  \gamma^K(t,x)=\sum_{k=-\frac K2}^{\frac K2-1}\hat \gamma^K_k(t)e^{\ii \frac kLx}
\end{align}
and the approximation of the Cooper pair density reads
\begin{align}\label{eqn-Int-alpha-colloc}
  \alpha^K(t,x)=\sum_{k=-\frac K2}^{\frac K2-1}\hat \alpha^K_k(t)e^{\ii \frac kLx}.
\end{align}
Inserting this approximations into the infinite dimensional BCS equations~\eqref{eqn-Int-BCS-k} yields
a finite dimensional system of ordinary differential equations (ODEs).
\subsection{System of ordinary differential equations}

The system of ordinary differential equations we end up with after applying the Fourier collocation is given by
\begin{align}
 &\ii\dot\gamma_k(t)=\alpha_k(t)\overline{\left(\hat V\ast\alpha\right)}_k
 -\overline{\alpha_k(t)}\left(\hat V\ast\alpha\right)_k,\label{eqn-Int-dot-gamma}\\
 &\ii\dot\alpha_k(t)=2\left(\frac{k^2}{L^2}-\mu\right)\alpha_k(t)-
 \left(2\gamma_k(t)-1\right)\left(\hat V\ast\alpha\right)_k,\label{eqn-Int-dot-alpha}\\
 &-\frac K2\le k\le\frac K2-1\nonumber,
\end{align}
where, for the sake
of readability, we have replaced $\hat\gamma^K$ and $\hat\alpha^K$ by $\gamma$ and $\alpha$, respectively.

\subsection{System for a contact interaction}

For a contact interaction $V(x)=-a\delta(x)$, $a>0$, which is the most popular interaction model in
physics, we have
\begin{align}
 \hat V(k)=-\frac a{2L\pi}, \quad  -\frac K2\le k\le \frac K2-1.
\end{align}
Hence, the convolution term in the self-consistent Hamiltonian on the $K$ dimensional subspace is given by
\begin{align}\label{eqn-Int-conv-delta}
 \left(\hat V\ast\hat\alpha^K\right)_k(t)=-\frac a{2L\pi}\sum_{j=-\frac K2}^{\frac K2-1}\hat\alpha^K_j(t).
\end{align}
With this relation, the equations of motion become
\begin{align}
 &\ii\dot\gamma_k(t)=-\frac a{L\pi}\left(\alpha_k(t)\overline{\sum_{j=-\nicefrac K2}^{\nicefrac K2-1}\alpha_j(t)}
 -\overline{\alpha_k(t)}\sum_{j=-\nicefrac K2}^{\nicefrac K2-1}\alpha_j(t)\right),\label{eqn-Int-dot-gamma-delta}\\
 &\ii\dot\alpha_k(t)=2\left(\frac{k^2}{L^2}-\mu\right)\alpha_k(t)+\frac a{L\pi}\sum_{j=-\nicefrac K2}^{\nicefrac K2-1}\alpha_j(t)
 \left(2\gamma_k(t)-1\right),\label{eqn-Int-dot-alpha-delta}\\
 &-\frac K2\le k\le\frac K2-1\nonumber.
\end{align}
With
\begin{align}
 p_k(t):=\operatorname{Re}\alpha_k(t),\\
 q_k(t):=\operatorname{Im}\alpha_k(t),
\end{align}
we can rewrite the equation of motion for $\gamma_k(t)$ as
\begin{align}\label{eqn-Int-dot-gamma-real}
 &\dot\gamma_k(t)=\frac{2a}{L\pi}\left(q_k(t)\sum_{j=-\nicefrac K2}^{\nicefrac K2-1}p_j(t)
 -p_k(t)\sum_{j=-\nicefrac K2}^{\nicefrac K2-1}q_j(t)\right).
\end{align}
From this expression we can see very easily
that $\gamma_k(t)$ is a real quantity whenever $\gamma_t(0)$ is so. As $\gamma$ represents the physical particle density, which
is real by definition, we can safely assume $\gamma_k(t)$ to be real in the following.
\subsection{Constants of motion}
For later use we mention that the coupled system~\eqref{eqn-Int-dot-gamma},\eqref{eqn-Int-dot-alpha} possesses some
important constants of motion:
\begin{itemize}
 \item 
It can readily be seen that the matrix $H_{\Gamma(t)}$ in the BCS equations~\eqref{eqn-Int-BCS-k} is self-adjoint. Together with the 
commutator structure of the equations of motion~\eqref{eqn-Int-BCS-k}, this implies that
the evolution of $\Gamma(t)$ is unitary. Consequently, its eigenvalues are preserved along the evolution. A little bit of algebra
shows that these eigenvalues are given by
\begin{align}
 \lambda^{\pm}_k=\frac12\pm\sqrt{\left(\gamma_k(t)-\frac12\right)^2+|\alpha_k(t)|^2}.
\end{align}
\item The discretized analog of the free energy functional~\eqref{eqn-Int-free-energy} in the case of an interaction potential is given by
\begin{align}\label{eqn-Int-discrete-E}
  F^K(\gamma(t),\alpha(t))&:=\sum_{k=-\nicefrac K2}^{\nicefrac K2-1}\left(\frac{k^2}{L^2}-\mu\right)\gamma_k(t)\nonumber\\
  \phantom{F^K(\gamma(t),\alpha(t))=}&+\frac1{2\pi}\int_0^{2\pi}V(x)\left|\alpha(t,x)\right|^2,\nonumber\\
 \phantom{F^K(\gamma(t),\alpha(t))=}& +T\sum_{k=-\nicefrac K2}^{\nicefrac K2-1}[\lambda^{+}_k\log(\lambda^{+}_k)+\lambda^{-}_k\log(\lambda^{-}_k)],
\end{align}
and can be shown to be preserved, too.
\end{itemize}
\subsection{Numerical notation}\label{IntSec:Not}
From a numerical point of view, the coupled system~\eqref{eqn-Int-dot-gamma},\eqref{eqn-Int-dot-alpha}, when supplemented 
by some initial data, represents an initial value problem
\begin{align}\label{eqn-Int-def-initial-value-problem}
 \begin{cases}\dt{\mathbf y(t)}&=f(\mathbf y(t)),\\
         \mathbf y(0)&=\mathbf y_0,\end{cases}
\end{align}
with $\mathbf y\in\mathbb C^{2K}$. Formally, the aim of this paper is to find a numerical approximation to the exact
flow of such an initial value problem. For this, we denote a time step by $\tau$ and the flow over such a time, i.e., 
the smooth map between $\mathbf y(t)$ and $\mathbf y(t+\tau)$, by $\Phi_{\tau,f}(\mathbf y(t))$. Its numerical
approximation will be denoted by $\Phi^\text{num}_{\tau,f}$.

Both of the numerical flows we present in this work rely on the fast calculation of the convolutions appearing at
the right hand side of the equations of motion~\eqref{eqn-Int-dot-gamma},\eqref{eqn-Int-dot-alpha}. Let us
turn towards this now.
\section{Calculating the Convolution Terms}
We denote by $\mathcal F$ the Fourier transform of a vector of length $K=2^N$, $N\in\mathbb N$, and
by $\mathcal F^{-1}$ its inverse. With the help of the fast Fourier transform (FFT) algorithms, these
operations can be calculated efficiently in $\mathcal O(N\cdot K)$ operations, see, e.g.,~\cite[Chapter 12]{1992F&PT..8..NR}.

Furthermore, the convolution of two $K$ dimensional vectors $a$ and $b$ can be computed by
\begin{align}
 a\ast b=\mathcal F^{-1}\left(\left(\mathcal F a\right)\cdot \left(\mathcal F b\right)\right),
\end{align}
with $\cdot$ denoting pointwise multiplication. Taking this into account, we can efficiently calculate
the convolution terms as outlined in Fig.~\ref{fig-calcconv}. There, we have defined
\begin{align}
 V_j:=V\left(\nicefrac{L2\pi}K\cdot j\right), \quad j=-\nicefrac K2,...,\nicefrac K2-1.
\end{align}
\verbdef{\verbtime}{calc_convolution}
\begin{figure}[h!]
 \centering
 \small
 \begin{minipage}[t]{.98\linewidth}
 \begin{algorithm}[H]
\hspace{-.2cm}\BlankLine
\hspace{-.2cm}$\text{conv}=\text{FFT}(\alpha)$\\
\hspace{-.2cm}\For{$j = -\nicefrac K2$ \KwTo $\nicefrac K2-1$}{
\hspace{-.2cm}$\text{conv}_j=\text{conv}_j\cdot V_j$.\\
\hspace{-.2cm}}
\hspace{-.2cm}$\text{conv}=\text{invFFT}(\text{conv})$\\
\caption{\verbtime}
\end{algorithm}
\end{minipage}
\caption{Sketch of the algorithm calc\_convolution, which uses the FFT and its inverse to efficiently calculate the convolution
between $\hat\alpha$ and $\hat V$.}
\label{fig-calcconv}
 \end{figure}
 The algorithm only takes $\mathcal O(N\cdot K)$ operations. When considering a system with contact interaction, i.e., when integrating
 the evolution equations~\eqref{eqn-Int-dot-gamma-real},\eqref{eqn-Int-dot-alpha-delta}, the convolution terms are just a sum
 over the entries of the vector $\alpha$. Hence, the CPU effort in this case is only $\mathcal O(K)$.

\section{Nonlinear Splitting Integrator}\label{IntSec:Nonlinear}
BCSInt, the integrator we present in this Section, is based on the conservation of the eigenvalues of $\Gamma(t)$.
These eigenvalues being conserved, the following equality holds
\begin{align}\label{eqn-Int-equality}
 \left(\gamma_k(t)-\frac12\right)^2+|\alpha_k(t)|^2=\left(\gamma_k(0)-\frac12\right)^2+|\alpha_k(0)|^2.
\end{align}
With the help of this relation, we can eliminate $\gamma_k(t)$ in the equations of motion for $\alpha_k(t)$ as we show now.
\subsection{Decoupled system}
Solving Eq.~\eqref{eqn-Int-equality} for $\gamma_k$ yields
\begin{align}\label{eqn-Int-gamma-of-alpha}
  \gamma_k(t)=\frac12\pm\sqrt{h(k)-|\alpha_k(t)|^2},
\end{align}
with the auxiliary function
\begin{align}
  h(k):=\left(\gamma_k(0)-\frac12\right)^2+|\alpha_k(0)|^2.
\end{align}
The sign in relation~\eqref{eqn-Int-gamma-of-alpha} can usually be inferred from physical information. In
our study~\cite{hainzl2015comparing}, for example, the initial values had to be such that $\gamma_k(0)$ was
greater than $\nicefrac12$ for $\mu>\nicefrac{k^2}{L^2}$ and less than or equal to $\nicefrac12$ for $\mu\le \nicefrac{k^2}{L^2}$.

Inserting the just-derived expression~\eqref{eqn-Int-gamma-of-alpha} for $\gamma_k(t)$ into the equations of motion for $\alpha_k(t)$,
we get the nonlinear system 
\begin{align}\label{eqn-Int-dotalphahat-reduced}
 &\ii\dot\alpha_k(t)=2\left(\frac{k^2}{L^2}-\mu\right)\alpha_k(t)\nonumber\\
 &\phantom{\ii\dot\alpha_k(t)}\pm\frac a{L\pi}
 \sqrt{h(k)-|\alpha_k(t)|^2}\sum_{j=-\nicefrac K2}^{\nicefrac K2-1}\alpha_t(j),\\
 &-\frac K2\le k\le \frac K2-1.\nonumber
\end{align}
Having decoupled the system, we can now turn towards its time evolution.
\subsection{BCSInt}
The nonlinear system~\eqref{eqn-Int-dotalphahat-reduced}, together with some suitable initial data, gives an initial value problem
\begin{align}\label{eqn-Int-def-initial-value-problem}
 \begin{cases}\ii\dt{\vec\alpha(t)}&=\tilde f(\vec\alpha(t)),\\
         \vec\alpha(0)&=\vec\alpha_0,\end{cases}
\end{align}
for
\begin{align}
  \vec\alpha=\begin{pmatrix}\alpha_{-\nicefrac K2}(t)&\hdots&\alpha_{\nicefrac K2-1}(t)\end{pmatrix}^T\in \mathbb C^K.
\end{align}
The right hand side of the differential equation can be written as the sum of two terms,
\begin{align}\label{eqn-Int-def-split}
\tilde f(\vec\alpha)=A\vec\alpha+f_1(\vec\alpha),
\end{align}
where $f_1$ represents the nonlinear term and where $A$ is the matrix
\begin{align}
 A=\operatorname{diag}\left(2\left(\frac{\left(-\frac K2\right)^2}{L^2}-\mu\right),...,2\left(\frac{\left(\frac K2-1\right)^2}{L^2}-\mu\right)\right).
\end{align}
This linear part resembles the kinetic part of the linear Schr\"odinger equation. Its flow $\Phi_{\tau,A}$ can be calculated exactly as
\begin{align}
 \Phi_{\tau,A}(\vec\alpha)=\operatorname{diag}\left(e^{-\ii2\left(\frac{(-K)^2}{4L^2}-\mu\right)\tau},...,
 e^{-\ii2\left(\frac{(K-2)^2}{4L^2}-\mu\right)\tau}\right)\vec\alpha.
\end{align}
With regard to $f_1$, it has a much smaller Lipschitz constant than the complete right hand side
$f$, wherefore $\Phi_{\tau,f_1}$ can be approximated by some standard
integration scheme. We than follow the idea of~\cite{strang1968construction} and set
\begin{align}
 \Phi^\text{num}_{\tau,\tilde f}(\vec\alpha(0))=\left(\Phi_{\nicefrac\tau2,A}\circ\Phi^\text{num}_{\tau,f_1}\circ\Phi_{\nicefrac\tau2,A}\right)(\vec\alpha(0)).
\end{align}
Applying this operation successively yields an approximation to the exact solution at times $t=n\tau$, $n=1,2,...$~. 
Its error decreases quadratically as a function of the step size $\tau$ as long as $\Phi^\text{num}_{\tau,f_1}$
is a second-or-higher order approximation to $\Phi_{\tau,f_1}$, see, e.g.~\cite[Chapter~II.5]{2006H&LW..8..679M}.

Just as every exact flow, $\Phi_{\tau,A}$ satisfies
\begin{align}
 \Phi_{t,A}\circ\Phi_{s,A}=\Phi_{t+s,A}.
\end{align}
Hence, when applying many time steps of the numerical scheme in a row, one can combine the last sub-step of the previous
step with the first sub-step of the next step, thus saving computational costs. 
We illustrate the resulting procedure in Fig.~\ref{fig-Int-BCSInt}.
\verbdef{\verbtime}{BCSInt}
\begin{figure}[h!]
 \centering
 \small
 \begin{minipage}[t]{.98\linewidth}
 \begin{algorithm}[H]
\hspace{-.2cm}\BlankLine
\hspace{-.2cm}$\vec\alpha=\Phi_{\nicefrac\tau2,A}(\vec\alpha_0)$\\
\hspace{-.2cm}\For{$n = 0$ \KwTo $N$}{
\hspace{-.2cm}$\vec\alpha=\Phi^\text{num}_{\tau,f_1}(\vec\alpha)$.\\
\hspace{-.2cm}$\vec\alpha=\Phi_{\tau,A}(\vec\alpha)$.\\
\hspace{-.2cm}}
\hspace{-.2cm}$\vec\alpha=\Phi_{-\nicefrac\tau2,A}(\vec\alpha)$\\
\caption{\verbtime}
\end{algorithm}
\end{minipage}
\caption{Sketch of our algorithm BCSInt which for a given initial value $\vec\alpha_0$ and a given step size $\tau$
approximates $\vec\alpha(N\tau)=\Phi_{N\tau,\tilde f}(\vec\alpha_0)$.}
\label{fig-Int-BCSInt}
 \end{figure}

Concerning $\Phi^\text{num}_{\tau,f_1}$,
in the study~\cite{hainzl2015comparing} it has been calculated via the fifth order explicit Cash--Karp Runge--Kutta scheme proposed in~\cite{1992F&PT..8..NR}.
In the experiment Section~\ref{IntSec:Exp} below, we will also test the second order explicit midpoint rule.
In this case, $\Phi^\text{num}_{\tau,f_1}$ is calculated as outlined in Fig.~\ref{fig-Int-calcPhif1}. 
\begin{figure}[h!]
\small
\centering
\begin{minipage}[t]{.49\linewidth}
\begin{algorithm}[H]
\hspace{-.2cm}\BlankLine
\hspace{-.2cm}$\dot{\mathbf Y}=$calc$f_1(\vec\alpha)$\\
\hspace{-.2cm}\For{$k=-\nicefrac K2$ \KwTo $\nicefrac K2-1$}{
\hspace{-.2cm}  $Y(k)=\alpha_k(t)+\tau/2\dot{\mathbf Y}(k)$\\
\hspace{-.2cm}}
\hspace{-.2cm}$\dot{\mathbf Y}=$calc$f_1(\mathbf Y)$\\
\hspace{-.2cm}\For{$k=-\nicefrac K2$ \KwTo $\nicefrac K2-1$}{
\hspace{-.2cm}  $\alpha_k(t)=\alpha_k(t)+\tau\dot{\mathbf Y}(k)$\\
\hspace{-.2cm}}
\caption{calc$\Phi_{f_1}$}
\end{algorithm}
\end{minipage}
\begin{minipage}[t]{.49\linewidth}
\begin{algorithm}[H]
\hspace{-.2cm}\BlankLine
\hspace{-.2cm}$c=\text{calc\_convolution}(V,\alpha)$\\
\hspace{-.2cm}\For{$k=-\nicefrac K2$ \KwTo $\nicefrac K2-1$}{
\hspace{-.2cm}  $d=\sqrt{h(k)-|\alpha_k(t)|^2}$\\
\hspace{-.2cm}  $f_1(k)=-\ii c\cdot d$\\
\hspace{-.2cm}}
\caption{calc$f_1$}
\end{algorithm}
\end{minipage}
 \caption{The left panel shows the algorithm which for 
 a given value $\vec\alpha(n\tau)$ and a given time step $\tau$ calculates $\vec\alpha((n+1)\tau)=\Phi^\text{num}_{\tau,f_1}(\vec\alpha(n\tau))$ with the explicit
 midpoint rule. The right panel shows the algorithm which for a given value $\vec\alpha$ calculates $f_1(\vec\alpha)$.}
 \label{fig-Int-calcPhif1}
\end{figure}
\subsection{Number of operations}\label{IntSubSec:NOOInt}
In order to analyze BCSInt's efficiency, we count the number of real operations which are executed per call of our 
implementations, which, to the best of our knowledge, have been implemented in the most efficient way possible.
We do not weight 
the costs of different operations, i.e., the square root in calc$f_1$, cf.~Fig.~\ref{fig-Int-calcPhif1}, also counts as a single operation. The number
of operations as a function of the number of basis functions $K$ 
for the various sub-algorithms and BCSInt as a whole are listed in Tab.~\ref{tab-Int-NumOp-BCSInt}. We mention that, if we substitute
the fifth order Cash--Karp scheme for the explicit midpoint rule in calc$\Phi_{f_1}$, the number of operations for calc$\Phi_{f_1}$
increases to $6\times\text{calc\_convolution}+38K$.
  \begin{table}[h!]
   \centering
  \scalebox{0.7}{
   \begin{tabular}{|c|c|}
    \hline
     Algorithm&$\#$Operations per call\\
    \hline
     Calculation of $\Phi_{\tau,A}$& $14\cdot K+14$\\
     Calc$f_1$& $1\times\text{calc\_convolution}+12\cdot K+20$\\
     calc$\Phi_{f_1}$& $2\times\text{calc$f_1$}+8\cdot K+9=2\times\text{calc\_convolution}+32\cdot K+49$\\
     BCSInt& $2\times\text{calc\_convolution}+46\cdot K+63$\\
     BCSInt for contact interaction& $58\cdot K+63$\\
    \hline
  \end{tabular}}\vspace{4mm}
  \caption{The required number of operations per step as a function of the dimension of the ODE system~\eqref{eqn-Int-dotalphahat-reduced}
  for the sub-algorithms of BCSInt and for BCSInt itself.}
  \label{tab-Int-NumOp-BCSInt}
 \end{table} 

Let us now introduce
our second integration scheme.

\section{Triple Splitting Integrator}\label{IntSec:FastInt}
For our second scheme, we consider the coupled system~\eqref{eqn-Int-dot-gamma},\eqref{eqn-Int-dot-alpha} as a whole. 
From a numerical perspective,
we have an initial value problem for
\begin{align}
 &\mathbf y(t)=\begin{pmatrix}\vec\gamma(t)\\\vec\alpha(t)\end{pmatrix}\in\mathbb C^{2K},\\
  &\vec\gamma(t)=\begin{pmatrix}\gamma_{-\nicefrac K2}(t)&\hdots&\gamma_{\nicefrac K2-1}(t)\end{pmatrix}^T\in \mathbb R^K,\\
  &\vec\alpha(t)=\begin{pmatrix}\alpha_{-\nicefrac K2}(t)&\hdots&\alpha_{\nicefrac K2-1}(t)\end{pmatrix}^T\in \mathbb C^K,\\
\end{align}
whose right hand side $f(\mathbf y)$ can be split into three parts,
\begin{align}
 f\left(\vec\gamma,\vec\alpha\right)=\tilde A\mathbf y+g(\vec\alpha)+h(\vec\gamma,\vec\alpha).
\end{align}
Here, $\tilde A\mathbf y$ is the first term of the equation of motion~\eqref{eqn-Int-dot-alpha} for $\alpha$, i.e., 
\begin{align}
 \tilde A\begin{pmatrix}\vec\gamma\\\vec\alpha\end{pmatrix}=\begin{pmatrix}\vec\gamma\\A\vec\alpha\end{pmatrix},
\end{align}
which means that it represents the same action on $\alpha$ as $A$ in the nonlinear case above. The function $g(\vec\alpha)$
represents the right hand side of the evolution equation for $\gamma$ and $h(\vec\gamma,\vec\alpha)$ is the second term of Eq.~\eqref{eqn-Int-dot-alpha}.

We will now show that we can efficiently calculate the flows for all three subproblems. The calculation of $\Phi_{\tau,\tilde A}$ is nothing
other than $\Phi_{\tau,A}$ acting on $\vec\alpha$ with $\vec\gamma$ held constant. We thus, in fact, only have to consider the other two subproblems.

\subsection{Calculating $\Phi_{\tau,g}$}
For the subsystem
\begin{align}\label{eqn-Int-subproblem-g}
 \begin{cases}\dt{\vec\gamma(t)}&=g(\vec\alpha(t)),\\
         \vec\gamma(0)&=\vec\gamma_0,\end{cases}
\end{align}
the right hand side does not depend on the quantity to be evolved. Therefore, the solution of the initial value problem~\eqref{eqn-Int-subproblem-g}
at time $t$ is trivially given by
\begin{align}
 \vec\gamma(t)=\Phi_{t,g}(\vec\gamma(0))=\vec\gamma(0)+t\cdot g(\vec\alpha(0)).
\end{align}
Bearing in mind the reformulation~\eqref{eqn-Int-dot-gamma-real}, we calculate a step of 
$\Phi_{\tau,g}$ with the algorithm illustrated in Fig.~\ref{fig-Int-calcPsig}. Please note that in the case $V(x)=-a\delta(x)$,
the convolution is replaced by the sum~\eqref{eqn-Int-conv-delta}. Hence, $\Phi_{\tau,g}$ can even be calculated in $\mathcal O(K)$
operations.
\begin{figure}[h!]
 \centering
 \small
 \begin{minipage}[t]{.98\linewidth}
 \begin{algorithm}[H]
\hspace{-.2cm}\BlankLine
\hspace{-.2cm}$p=2a/(L\pi)\sum_{k=-\frac K2}^{\frac K2-1}p_k$\\
\hspace{-.2cm}$q=2a/(L\pi)\sum_{k=-\frac K2}^{\frac K2-1}q_k$\\
\hspace{-.2cm}\For{$k=-\nicefrac K2$ \KwTo $\nicefrac K2-1$}{
\hspace{-.2cm}$\gamma_k(t)=\gamma_k(0)+\tau\cdot(q_k\cdot p-p_k\cdot q)$\\
\hspace{-.2cm}}
\caption{calc$\Phi_g$}
\end{algorithm}
\end{minipage}
\caption{Sketch of the algorithm which for given values $\vec\gamma(0)$, $\vec\alpha(0)$ and a given step size $\tau$
calculates the solution $\vec\gamma(\tau)=\Phi_{\tau,g}(\vec\gamma(0))$ to the initial value problem~\eqref{eqn-Int-subproblem-g}.}
\label{fig-Int-calcPsig}
 \end{figure}

\subsection{Calculating $\Phi_{\tau,h}$}
We consider the subproblem
\begin{align}\label{eqn-Int-subproblem-h}
 \begin{cases}\ii\dt{\vec\alpha(t)}&=h(\vec\gamma(0),\vec\alpha(t)),\\
         \vec\alpha(0)&=\vec\alpha_0.\end{cases}
\end{align}
As the right hand side's Lipschitz constant is small, we can apply a standard integration scheme. This yields
the algorithm outlined in Fig.~\ref{fig-Int-calcPhih1}.
\begin{figure}[h!]
\small
\centering
\begin{minipage}[t]{.49\linewidth}
\begin{algorithm}[H]
\hspace{-.2cm}\BlankLine
\hspace{-.2cm}$\dot{\mathbf Y}=$calc\_$h(\vec\alpha)$\\
\hspace{-.2cm}\For{$k=-\nicefrac K2$ \KwTo $\nicefrac K2-1$}{
\hspace{-.2cm}  $Y(k)=\alpha_k(t)+\tau/2\dot{\mathbf Y}(k)$\\
\hspace{-.2cm}}
\hspace{-.2cm}$\dot{\mathbf Y}=$calc\_$h(\mathbf Y)$\\
\hspace{-.2cm}\For{$k=-\nicefrac K2$ \KwTo $\nicefrac K2-1$}{
\hspace{-.2cm}  $\alpha_k(t)=\alpha_k(t)+\tau\dot{\mathbf Y}(k)$\\
\hspace{-.2cm}}
\caption{calc$\Phi_{h}$}
\end{algorithm}
\end{minipage}
\begin{minipage}[t]{.49\linewidth}
\begin{algorithm}[H]
\hspace{-.2cm}\BlankLine
\hspace{-.2cm}$c=\text{calc\_convolution}(V,\vec\alpha)$\\
\hspace{-.2cm}\For{$k=-\nicefrac K2$ \KwTo $\nicefrac K2-1$}{
\hspace{-.2cm}  $d=2\cdot\gamma_k-1$\\
\hspace{-.2cm}  $h(k)=-\ii c\cdot d$\\
\hspace{-.2cm}}
\caption{calc\_$h$}
\end{algorithm}
\end{minipage}
 \caption{The left panel shows the algorithm which for 
 a given value $\vec\alpha(n\tau)$ and a given time step $\tau$ calculates $\vec\alpha((n+1)\tau)=\Phi^\text{num}_{\tau,h}(\vec\alpha(n\tau))$ with the explicit
 midpoint rule. The right panel shows the algorithm which for a given value $\vec\alpha$ calculates $h(\vec\gamma,\vec\alpha)$.}
 \label{fig-Int-calcPhih1}
\end{figure}
The appealing fact about our triple splitting is that in the case of a contact interaction, the subproblem~\eqref{eqn-Int-subproblem-h}
can be solved exactly in $\mathcal O(k)$ operations as we show now.

Introducing $\vec b\in\mathbb R^K$ via
\begin{align}
 b_k=\frac a{L\pi}\left(2\gamma_k(0)-1\right),
\end{align}
and the $K\times K$-matrix
\begin{align}
 B=\underbrace{\begin{pmatrix}b_{-\nicefrac K2}&\hdots&b_{-\nicefrac K2}\\\vdots&\ddots&\vdots\\b_{\nicefrac K2-1}&\hdots&b_{\nicefrac K2-1}\end{pmatrix}}_K,
\end{align}
we can write 
\begin{align}
 h(\vec\gamma(0),\vec\alpha(t))=B\vec\alpha(t).
\end{align}
Hence, the solution to the initial value problem~\eqref{eqn-Int-subproblem-h} is given by
\begin{align}\label{eqn-Int-solution-ODE-alpha}
  \vec\alpha(\tau)=\Phi_{\tau,h}(\vec\alpha(0))=e^{-\ii B\tau}\vec\alpha(0)=:e^{\tilde B}\vec\alpha(0).
\end{align}
We now show that for a given $\vec\alpha(0)$, $\vec\alpha(\tau)$ can be calculated
in $\mathcal O(K)$ operations.

For a given $n\in\mathbb N$, we have
\begin{align}
  \tilde B^n=\underbrace{\begin{pmatrix}-\ii b_{-\nicefrac K2}\tau c^{n-1}&\hdots&-\ii b_{-\nicefrac K2}\tau c^{n-1}
  \\\vdots&\ddots&\vdots\\-\ii b_{\nicefrac K2-1}\tau c^{n-1}&\hdots&-\ii b_{\nicefrac K2-1}\tau c^{n-1}\end{pmatrix}}_K,
\end{align}
with
\begin{align}
  c=-\ii\tau\sum_{j=-\frac K2}^{\frac K2-1}b_j.
\end{align}
Consequently, with $\operatorname{Id}$ denoting the $K\times K$ identity matrix,
\begin{widetext}we have
\begin{align}
  \exp(-\ii\tau B)&=\operatorname{Id}+\sum_{n=1}^\infty\frac1{n!}\begin{pmatrix}-\ii b_{-\nicefrac K2}\tau c^{n-1}&\hdots&-\ii b_{-\nicefrac K2}\tau c^{n-1}
  \\\vdots&\ddots&\vdots\\-\ii b_{\nicefrac K2-1}\tau c^{n-1}&\hdots&-\ii b_{\nicefrac K2-1}\tau c^{n-1}\end{pmatrix}\\
  &=\operatorname{Id}+\frac1c\begin{pmatrix}-\ii b_{-\nicefrac K2}\tau(\exp(c)-1)&\hdots&-\ii b_{-\nicefrac K2}\tau(\exp(c)-1)
  \\\vdots&\ddots&\vdots\\-\ii b_{\nicefrac K2-1}\tau(\exp(c)-1)&\hdots&-\ii b_{\nicefrac K2-1}\tau(\exp(c)-1)\end{pmatrix}.
\end{align}
With this, the matrix-vector multiplication in Eq.~\eqref{eqn-Int-solution-ODE-alpha} yields
\begin{align}
  \exp(-\ii\tau B)\vec\alpha(0)&=\vec\alpha(0)-\frac{\ii\tau}c\begin{pmatrix}b_{-\nicefrac K2}(\exp(c)-1)\sum_{j=-\frac K2}^{\frac K2-1}\alpha_j(0)\\
  \vdots\\b_{\nicefrac K2-1}(\exp(c)-1)\sum_{j=-\frac K2}^{\frac K2-1}\alpha_j(0)\end{pmatrix}.
\end{align}
\end{widetext}
Thus, the solution of the initial value problem~\eqref{eqn-Int-subproblem-h} can efficiently be calculated
by the algorithm illustrated in Fig.~\ref{fig-Int-calcPsih}.
\begin{figure}[h!]
 \centering
 \small
 \begin{minipage}[t]{.98\linewidth}
 \begin{algorithm}[H]
\hspace{-.2cm}\BlankLine
\hspace{-.2cm}$c=-\ii\tau\sum_{k=-\frac K2}^{\frac K2-1}b_k$\\
\hspace{-.2cm}$s=\sum_{k=-\frac K2}^{\frac K2-1}\alpha_k(0)$\\
\hspace{-.2cm}$e=\exp(c)-1$\\
\hspace{-.2cm}\For{$k=-\nicefrac K2$ \KwTo $\nicefrac K2-1$}{
\hspace{-.2cm}$\alpha_k(t)=\alpha_k(0)-\ii\tau\cdot e\cdot s\cdot b_k/c$\\
\hspace{-.2cm}}
\caption{calc$\Phi_h$}
\end{algorithm}
\end{minipage}
\caption{Sketch of the algorithm which for given values $\vec\gamma(0)$, $\vec\alpha(0)$ and a given step size $\tau$
calculates the solution $\vec\alpha(\tau)=\Phi_{\tau,h}(\vec\alpha(0))$ to the initial value problem~\eqref{eqn-Int-subproblem-h}
for the case of a contact interaction.}
\label{fig-Int-calcPsih}
 \end{figure}

 Having found efficient algorithms for all three subproblems we have split the system into, we can now recompose them.
 \subsection{SplitBCS}
 As all the three flows $\Phi_{\tau,A}$, $\Phi_{\tau,g}$, and $\Phi_{\tau,h}$ are at least of second order, each symmetric composition
 of them gives rise to a second order integration scheme, see, e.g.~\cite[Chapter~II.5]{2006H&LW..8..679M}. We propose the composition
 \begin{align}\label{eqn-Int-fastBCS}
  \Phi^\text{num}_{\tau,f}=\Phi_{\tau,AghgA}:=\Phi_{\nicefrac\tau2,A}\circ\Phi_{\nicefrac\tau2,g}\circ\Phi_{\tau,h}
  \circ\Phi_{\nicefrac\tau2,g}\circ\Phi_{\nicefrac\tau2,A},
 \end{align}
 as this yields the fastest and most accurate scheme among the possible combinations as we will see in the next Section. If 
 even more accuracy were required, we could use a suitable composition of the scheme~\eqref{eqn-Int-fastBCS}; see~\cite{suzuki1990fractal,yoshida1990construction}
 for more information on compositions.
 
 In the same way as for the algorithm of Section~\ref{IntSec:Nonlinear},
 the last sub-step of each step can be combined with the first sub-step of the following step which reduces
 the CPU effort. Even more computational costs can be saved by paying heed to the following points.
 \begin{itemize}
  \item From Eq.~\eqref{eqn-Int-dot-gamma-real} it can be deduced that
  \begin{align}
   \frac{\mathrm d}{\mathrm dt}\left(\sum_{j=-\frac K2}^{\frac K2-1}\gamma_k(t)\right)=0.
  \end{align}
  Thus, the sum over all $\gamma_k(t)$, and, as a consequence, also the quantities $c$ and $e$ appearing in
  the calculation of $\Phi_{\tau,h}$, cf.~Fig.~\ref{fig-Int-calcPsih},
  are preserved along evolutions of the equations of motion. Hence, $c$ and $e$ only need to be calculated once
  at the start of the simulation when considering a contact interaction.
  \item  Both $\Phi_{\tau,g}$ and $\Phi_{\tau,h}$
  require the computation of the convolution, cf.~Figs.~\ref{fig-Int-calcPsig}~and~\ref{fig-Int-calcPsih}.
  However, $\Phi_{\tau,g}$ does not modify $\vec\alpha$ which means that the convolution in the first
  call of calc\_$h$
  is the same as the one already calculated in calc$\Phi_g$. 
  Hence, by suitably combining the calculation of
  \begin{align}\label{eqn-Int-Phighg}
   \Phi_{\tau,ghg}:=\Phi_{\nicefrac\tau2,g}\circ\Phi_{\tau,h}\circ\Phi_{\nicefrac\tau2,g}
  \end{align}
  into one algorithm, one can avoid redundancies\footnote{An efficient implementation of $\Phi_{\tau,ghg}$
  in c++ can be found on the author's homepage \href{http://na.uni-tuebingen.de/~seyrich/}{http://na.uni-tuebingen.de/$\sim$seyrich/}.}.
  \item The calculation of $\Phi_{\tau,A}$ can be made more efficient for both BCSInt and SplitBCS when a fixed step size is used.
  In this case, during each call of $\Phi_{\tau,A}$, $\cos$ and $\sin$ of $2\left(\nicefrac{k^2}{L^2}-\mu\right)\tau$, $k=\nicefrac K2,...,
  \nicefrac K2-1$, have to be calculated. But, if storage is not a problem, one only has to calculate the $\cos$ and $\sin$ once at the 
  beginning of the simulation as the arguments are the same in each step. This is what we did in our implementations. Accordingly, the number of operations specified in 
  Tabs.~\ref{tab-Int-NumOp-BCSInt}~and~\ref{tab-Int-NumOp-SplitBCS}, refer to this efficient version.
 \end{itemize}
 Putting everything together, we obtain our integrator SplitBCS as outlined in Fig.~\ref{fig-Int-SplitBCS}.
\verbdef{\verbtime}{SplitBCS}
\begin{figure}[h!]
 \centering
 \small
 \begin{minipage}[t]{.98\linewidth}
 \begin{algorithm}[H]
\hspace{-.2cm}\BlankLine
\hspace{-.2cm}$\vec\alpha=\Phi_{\nicefrac\tau2,A}(\vec\alpha(0))$\\
\hspace{-.2cm}\For{$n = 0$ \KwTo $N$}{
\hspace{-.2cm}$\left(\vec\gamma,\vec\alpha\right)=\Phi_{\tau,ghg}(\vec\gamma,\vec\alpha)$.\\
\hspace{-.2cm}$\vec\alpha=\Phi_{\tau,A}(\vec\alpha)$.\\
\hspace{-.2cm}}
\hspace{-.2cm}$\vec\alpha=\Phi_{-\nicefrac\tau2,A}(\vec\alpha)$\\
\caption{\verbtime}
\end{algorithm}
\end{minipage}
\caption{Sketch of our algorithm SplitBCS which for given initial values $\vec\gamma(0)$, $\vec\alpha(0)$ and a given step size $\tau$
approximates $\left(\vec\gamma(N\tau),\vec\alpha(N\tau)\right)^T=\Phi_{N\tau,f}(\vec\gamma(0),\vec\alpha(0))$.}
\label{fig-Int-SplitBCS}
 \end{figure}
 \subsection{Number of operations}\label{IntSubSec:NOOFast}
 In order to compare the efficiency of SplitBCS to the one of BCSInt, we count the number of operations required for the respective
 sub-algorithms and for SplitBCS as a whole, too. The result can be found in Tab.~\ref{tab-Int-NumOp-SplitBCS}.
 If the explicit midpoint rule is replaced by the Cash--Karp scheme, the number of operations in the calculation
 of $\Phi_{\tau,h}$ increases as for BCSInt.
  \begin{table}[h!]
   \centering
  \scalebox{0.6}{
   \begin{tabular}{|c|c|}
    \hline
     Algorithm&$\#$Operations per call\\
    \hline
     Calculation of $\Phi_{\tau,A}$& $14\cdot K+14$\\
     Calc\_$h$& $1\times\text{calc\_convolution}+7\cdot K+12$\\
     calc$\Phi_{f_1}$& $2\times\text{calc\_$h$}+8\cdot K+9=2\times\text{calc\_convolution}+22\cdot K+33$\\
     calc$\Phi_g$& $1\times\text{calc\_convolution}+6\cdot K+12$\\
     Calculation of $\Phi_{\tau,ghg}$ for contact interaction& $18\cdot K+39$\\
     SplitBCS& $3\times\text{calc\_convolution}+34\cdot K+53$\\
     SplitBCS for contact interaction& $32\cdot K+53$\\
    \hline
  \end{tabular}}\vspace{4mm}
  \caption{The required number of operations per step as a function of dimension of the ODE system~\eqref{eqn-Int-dot-gamma},\eqref{eqn-Int-dot-alpha}
  for the sub-algorithms of SplitBCS and for SplitBCS itself.}
  \label{tab-Int-NumOp-SplitBCS}
 \end{table} 
 We see that SplitBCS calculates one convolution more than BCSInt. Thus, BCSInt is expected to be faster
 for general settings with a huge number of basis functions.
 For the important case of a contact interaction, however, we are able to calculate $\Phi_{\tau,ghg}$ very efficiently. This
 is why we choose the composition~\eqref{eqn-Int-fastBCS} over other possible sequences of the subflows. With this, SplitBCS is even
 faster than BCSInt in the physically important setting.
 
 Let us now subject the schemes to numerical tests.
\section{Numerical Experiments}\label{IntSec:Exp}
All the numerical experiments presented here were run on a Core 2 Duo E6600 machine with $2.4$GHz and $4$GB RAM. In order to have physically
realistic data to start our experiments with, we chose a system which is slightly superconducting. Such a system
can be obtained by setting
\begin{align}
  &\hat\gamma_k(0)=\frac12-\frac{\nicefrac{k^2}{L^2}-\mu}2\frac{\tanh\left(\frac{\sqrt{(\nicefrac{k^2}{L^2}-\mu)^2+h^2}}{2T}\right)}{\sqrt{(\nicefrac{k^2}{L^2}-\mu)^2+h^2}}\\
  &\hat\alpha_k(0)=\frac{h}2\frac{\tanh\left(\frac{\sqrt{(\nicefrac{k^2}{L^2}-\mu)^2+h^2}}{2T}\right)}{\sqrt{(\nicefrac{k^2}{L^2}-\mu)^2+h^2}},
\end{align}
where $h=0.1$ is a small parameter. The critical temperature $T$ of the system depends on the chemical potential $\mu$ and on the interaction potential $V$. 
In the simulations presented here, we considered a system with a contact interaction $V(x)=-a\delta(x)$.
In this case, $T$ can
be calculated from the implicit formula, cf.~\cite{hainzl2015comparing},
\begin{align}\label{eqn-Int-Tc-integral}
 \frac{2\pi}a=\int_{\R}\frac{\tanh\left(\frac{p^2-\mu}{2T}\right)}{p^2-\mu}\mathrm dp.
\end{align}
For our simulations, we chose $a=\mu=1$ which yields $T=0.19$. 

As a measure of an integrator's accuracy, we used the discrete energy~\eqref{eqn-Int-discrete-E} which
is conserved along the exact solution of the ODE system~\eqref{eqn-Int-dot-gamma},\eqref{eqn-Int-dot-alpha}. 
Thus, the reliability
of a numerical integration scheme can be checked by tracking the relative error $\Delta F^K$, defined by
\begin{align}
  \Delta F^K(t)=\left|\frac{F^K(\vec\gamma(t),\vec\alpha(t))-F^K(\vec\gamma(0),\vec\alpha(0))}{F^K(\vec\gamma(0),\vec\alpha(0))}\right|,
\end{align}
along the numerical evolution.

\begin{figure} [htp]
  \centering
  \includegraphics[width=0.45\textwidth]{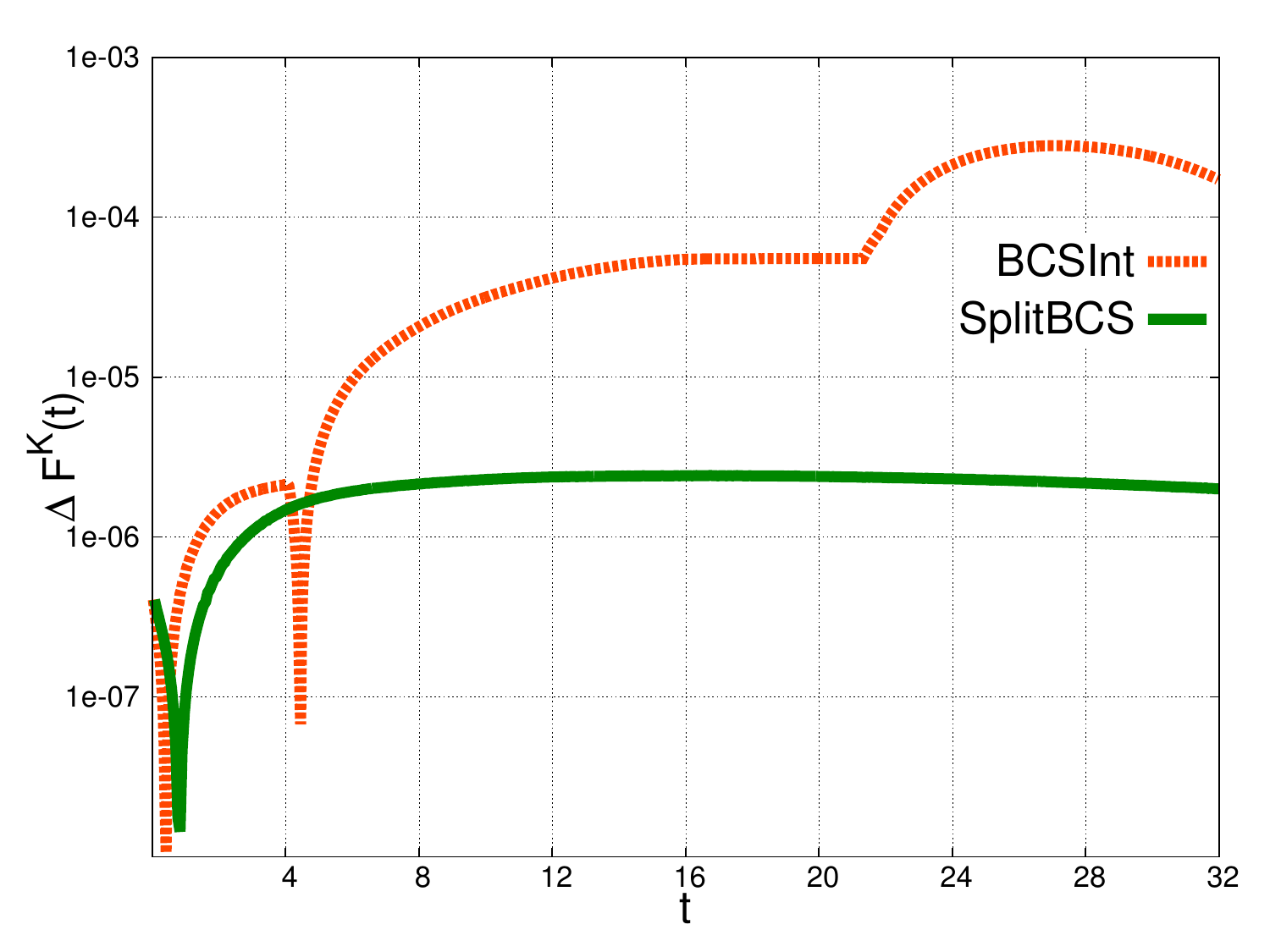}
  \includegraphics[width=0.45\textwidth]{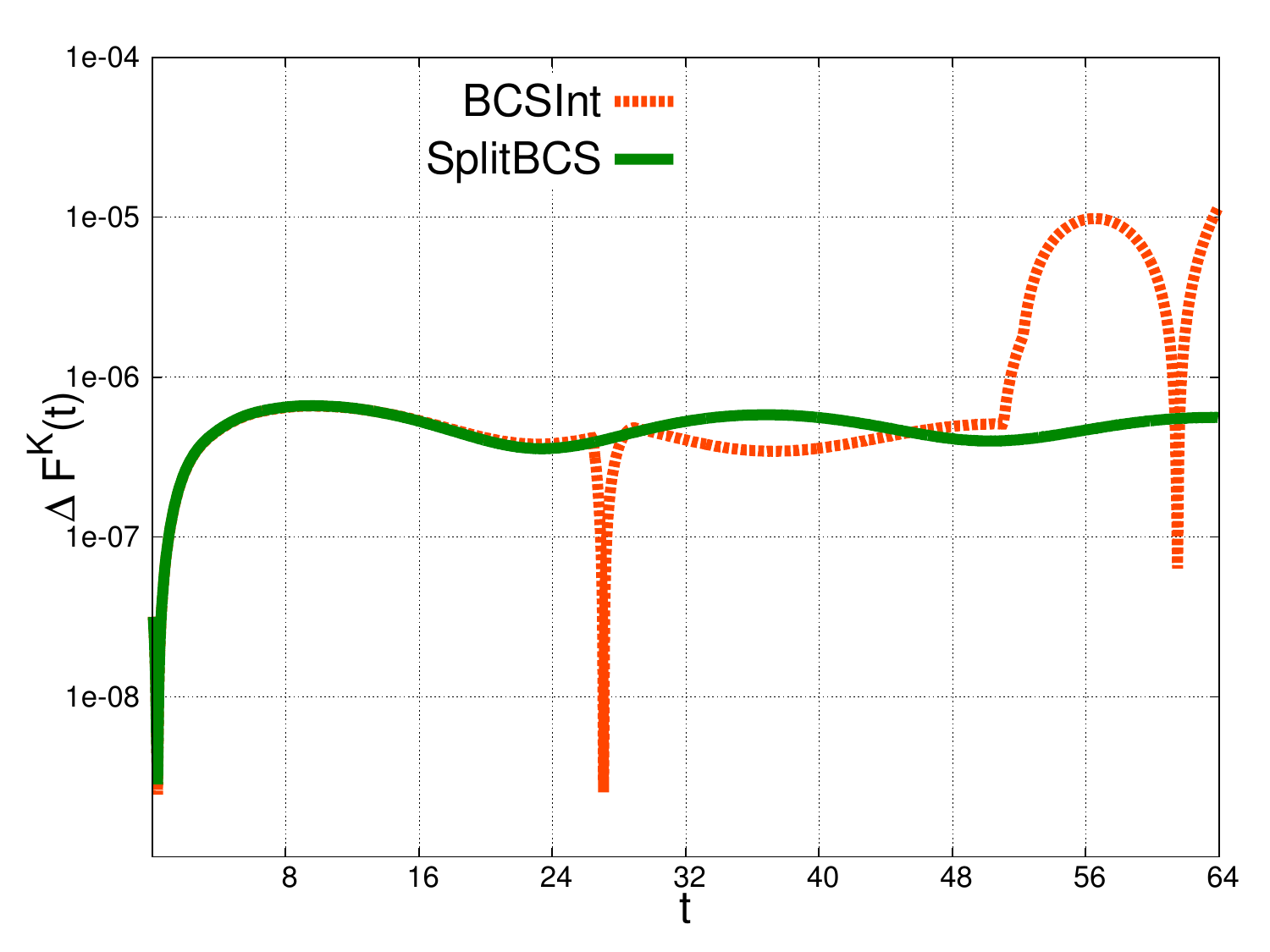}
  \caption{The relative error $\Delta F^K$ of the free energy as a function of integration time $t$ for SplitBCS and BCSInt in semilogarithmic scale. The left panel shows the result for $L=32$, 
  the right panel depicts the corresponding result for $L=64$.}
  \label{fig-comp-SplitBCS-BCSInt}
 \end{figure}

We first used this tool to compare SplitBCS to BCSInt with $\Phi_{\tau,f_1}$ calculated
via the fifth order Cash--Karp method. For this, we fixed $L=32$, $K=256\cdot L$ and chose a step size $\tau=\nicefrac{0.1}K$.
We evolved the system until $t=\mathcal O(L)$ with both integrators and plotted the relative error in the energy, $\Delta F^K$, against integration 
time $t$ in the left panel of Fig.~\ref{fig-comp-SplitBCS-BCSInt}. We repeated the procedure for $L=64$ and plotted the result in the right panel 
of Fig.~\ref{fig-comp-SplitBCS-BCSInt}.
Although the error increases slightly at the end of the integration for BCSInt, both schemes seem to be very accurate.
When comparing BCSInt with $\Phi_{\tau,f_1}$ calculated via the fifth-order Cash--Karp method to BCSInt where $\Phi_{\tau,f_1}$
was calculated with the explicit midpoint rule, we found no differences in the relative error of the energy. Hence, we recommend the use
of the latter method as it is much faster.

\begin{figure} [htp]
  \centering
  \includegraphics[width=0.45\textwidth]{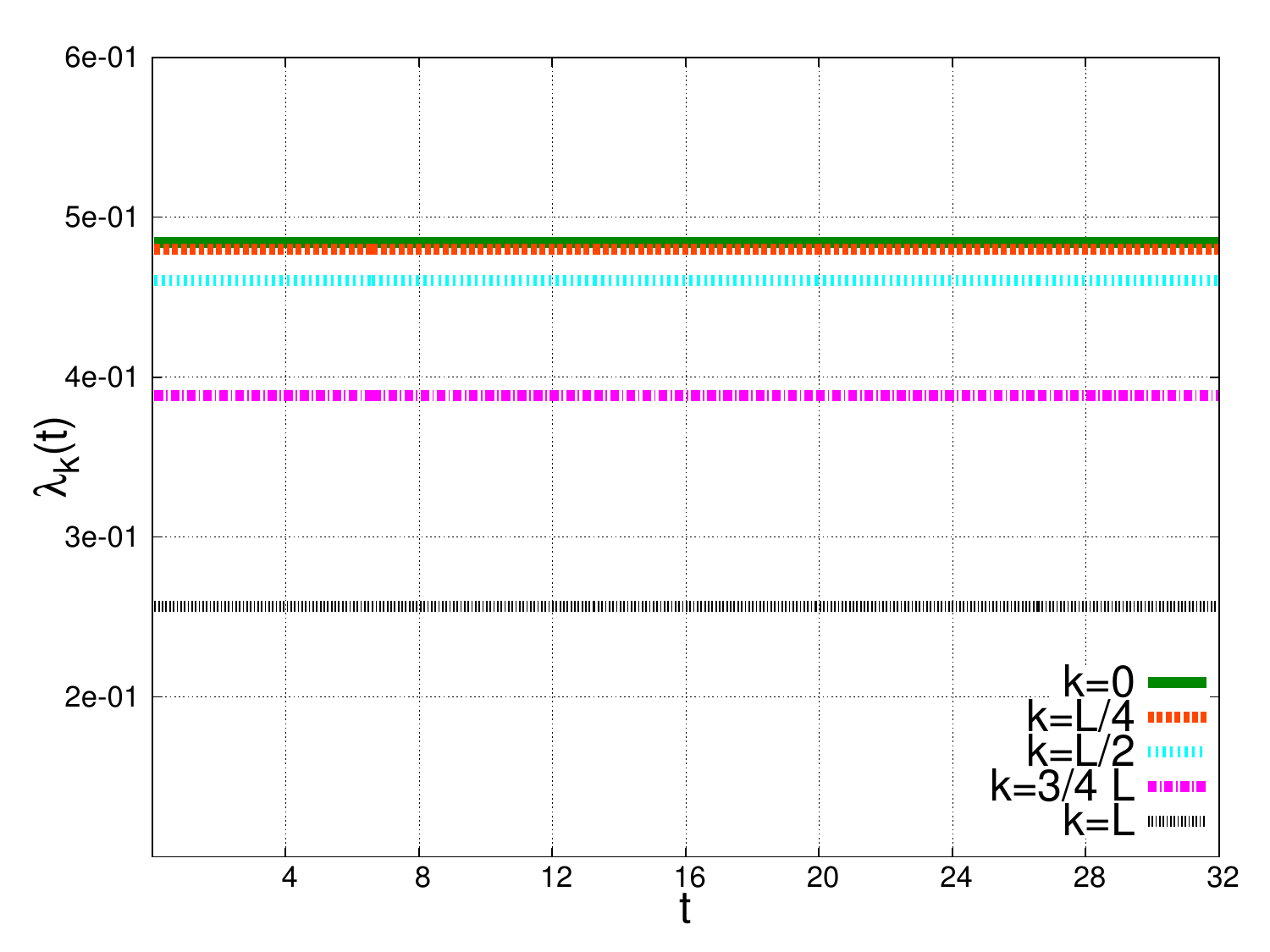}
  \includegraphics[width=0.45\textwidth]{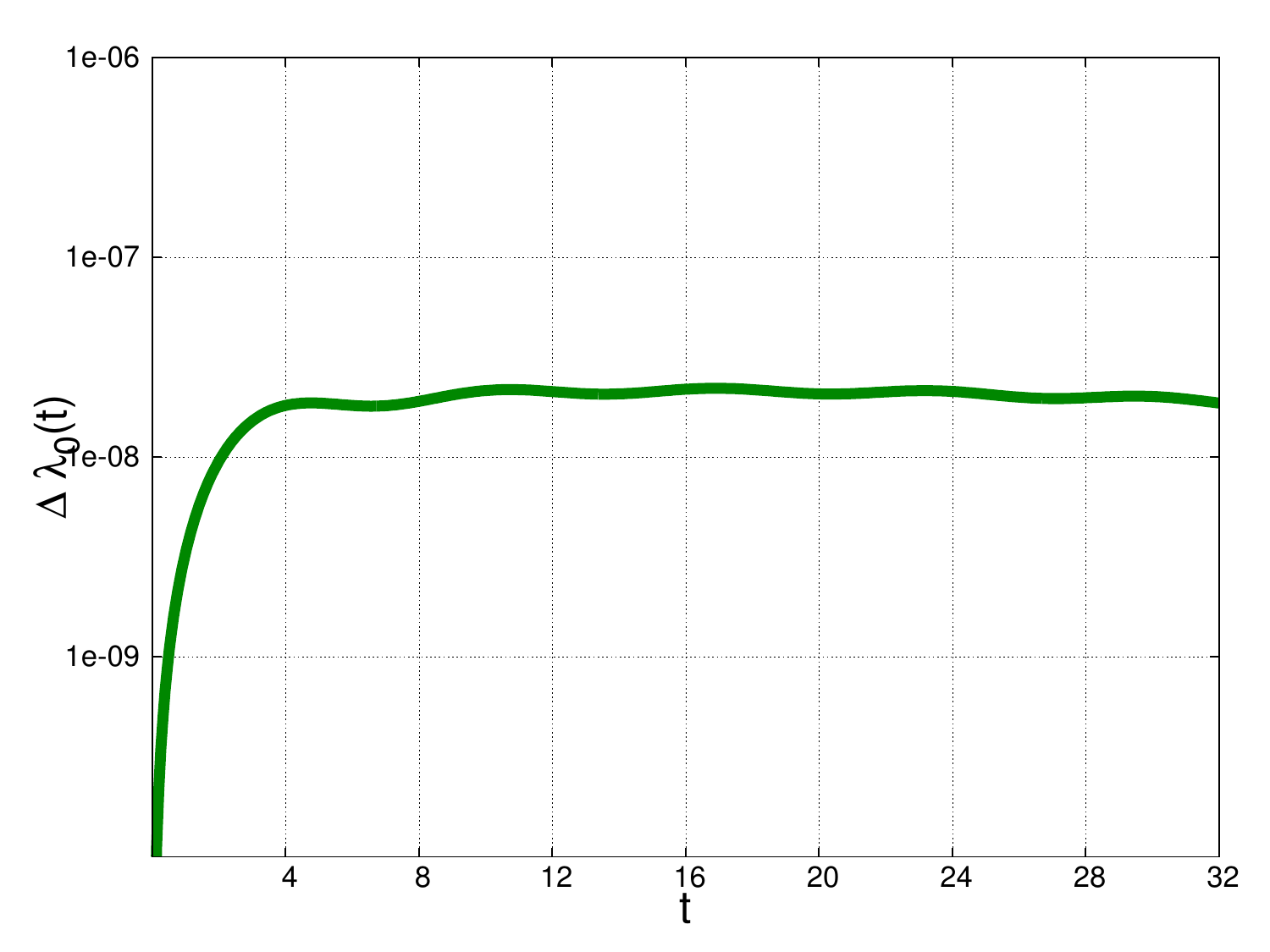}
  \caption{The left panel shows some eigenvalues $\lambda_k$ of the density matrix as a function of integration time $t$ for SplitBCS 
  applied to the system with $L=32$. 
  The right panel shows the corresponding relative error $\Delta\lambda_0$ of the density matrix' first eigenvalue
  in semilogarithmic scale.}
  \label{fig-Int-deltalambda}
 \end{figure}

Other physically relavant constants of motion are the eigenvalues $\lambda_k$ of the particle
density matrix $\Gamma$. BCSInt preserves them by construction. In order to check their behavior when using SplitBCS, 
we also tracked the eigenvalues together with their corresponding relative error,
\begin{align}
  \Delta \lambda_k(t)=\left|\frac{\lambda_k(t)-\lambda_k(0)}{\lambda_k(0)}\right|,
\end{align}
along the evolution. We found out that, up to very small rounding errors, all eigenvalues were preserved for SplitBCS, too. As an illustration,
we plot some eigenvalues and the relative error in $\lambda_0$ in Fig.~\ref{fig-Int-deltalambda}.

\begin{figure} [htp]
  \centering
  \includegraphics[width=0.45\textwidth]{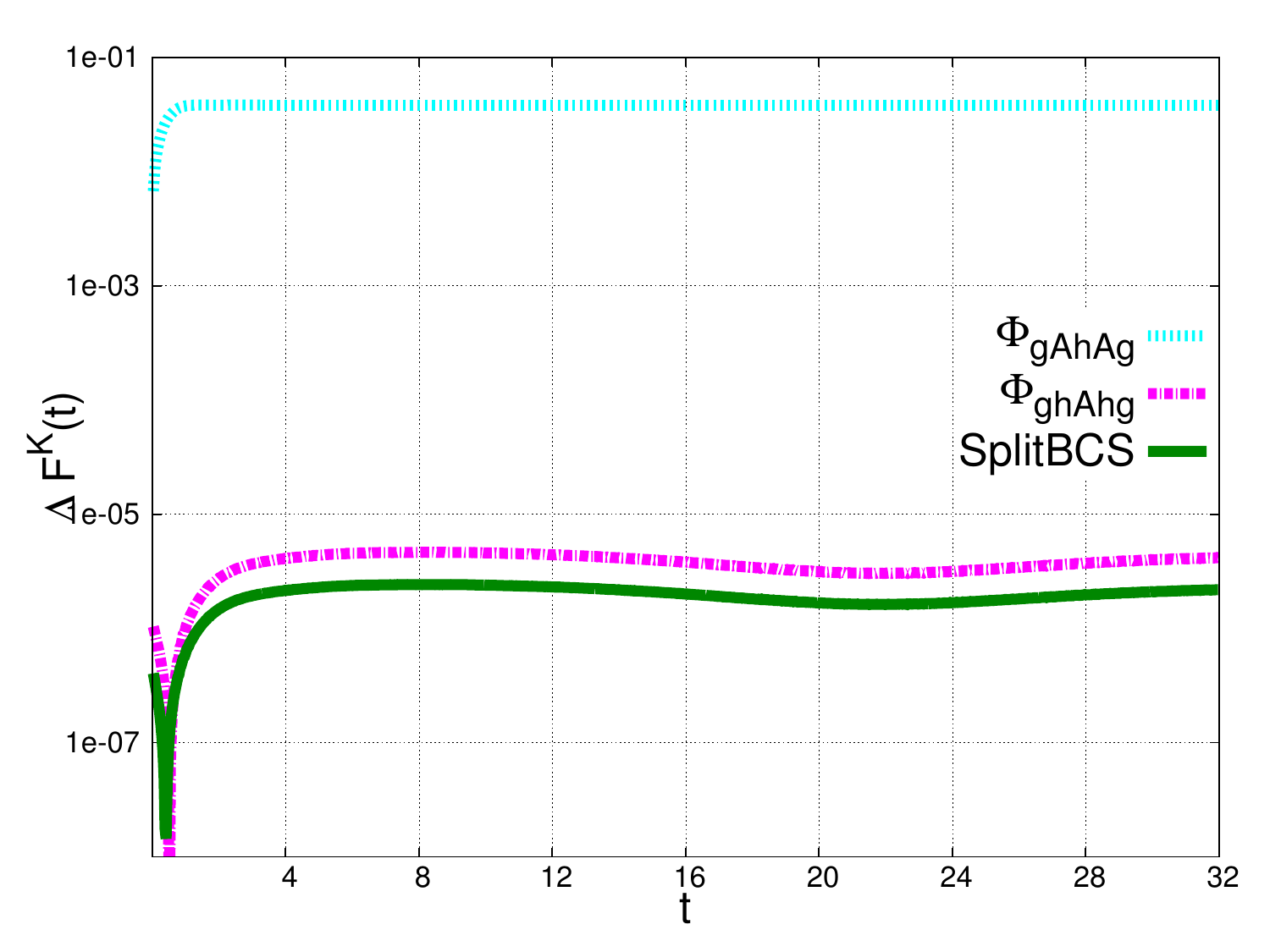}
  \includegraphics[width=0.45\textwidth]{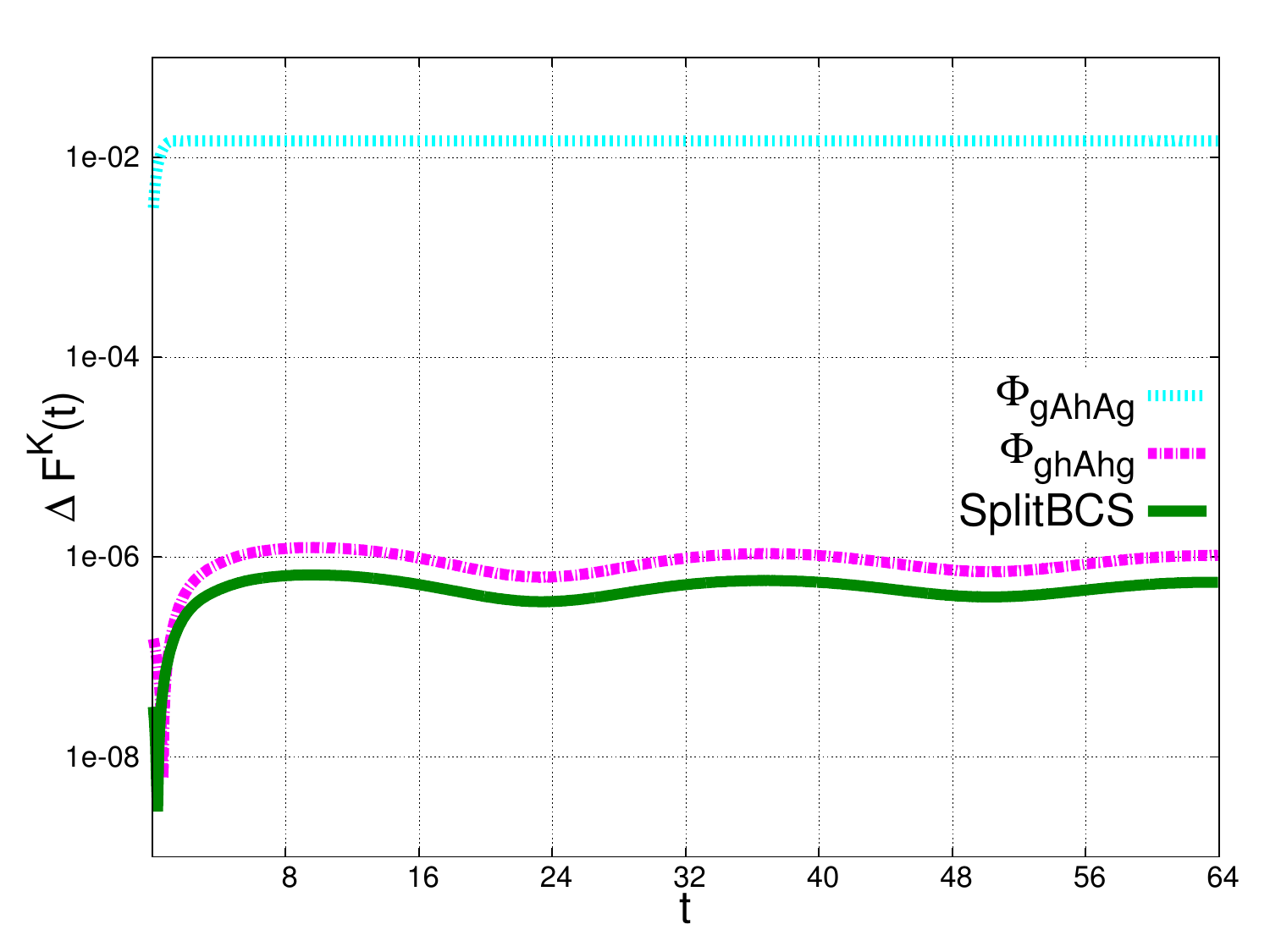}
  \caption{The relative error $\Delta F^K$ of the free energy as a function of integration time $t$ for SplitBCS and other possible compositions in semilogarithmic scale.
  The left panel shows the result for $L=32$, 
  the right panel depicts the corresponding result for $L=64$.}
  \label{fig-comp-SplitBCS-others}
 \end{figure}

With regard to SplitBCS, the question remains as to whether we could have done even better by choosing another sequence of the sub-flows than 
composition~\eqref{eqn-Int-fastBCS}. In order to go into this matter, we also evolved the systems for $L=32$ and $L=64$ for various
other compositions of the sub-flows $\Phi_{\tau,A}$,  $\Phi_{\tau,g}$ and  $\Phi_{\tau,h}$,
and again plotted $\Delta F^K$ as a function of the integration time $t$. The resulting plots
are shown in Fig.~\ref{fig-comp-SplitBCS-others}.
 We also tested the other possible sequences which are not shown in the plots. 
 However, we found out that the relative error in the energy seems only to depend on the spot of $\Phi_{\tau,A}$
 in the composition. This means that $\Phi_{\tau,AhghA}$ is as accurate as SplitBCS. But we could not find an equally efficient implementation for $\Phi_{\tau,hgh}$ as the one
 for $\Phi_{\tau,ghg}$. This is why we strongly recommend the use of the composition~\eqref{eqn-Int-fastBCS}, shortly SplitBCS, in simulations of the discrete BCS equations
 with a contact interaction.
 
\begin{figure} [htp]
  \centering
  \includegraphics[width=0.45\textwidth]{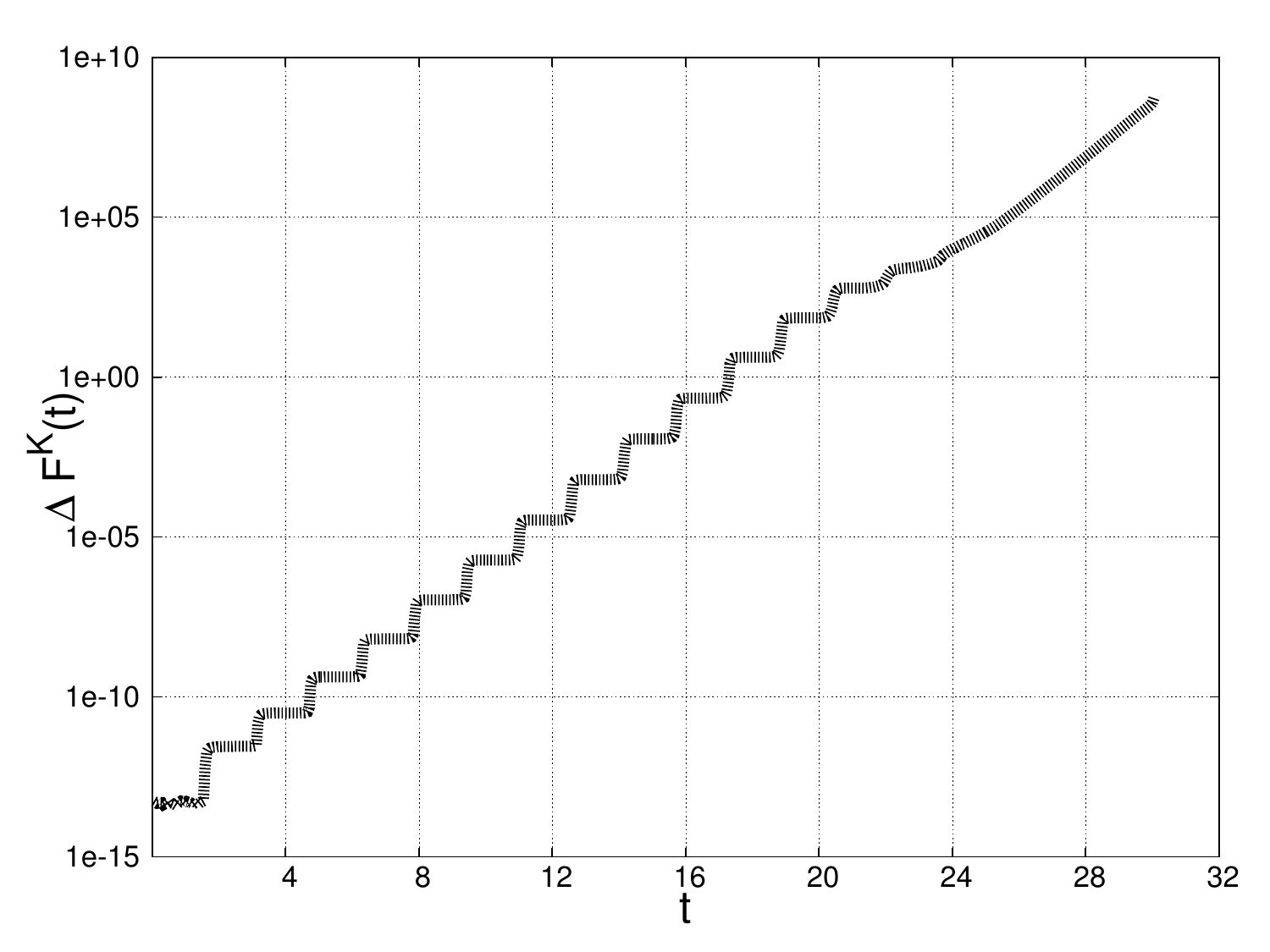}
  \includegraphics[width=0.45\textwidth]{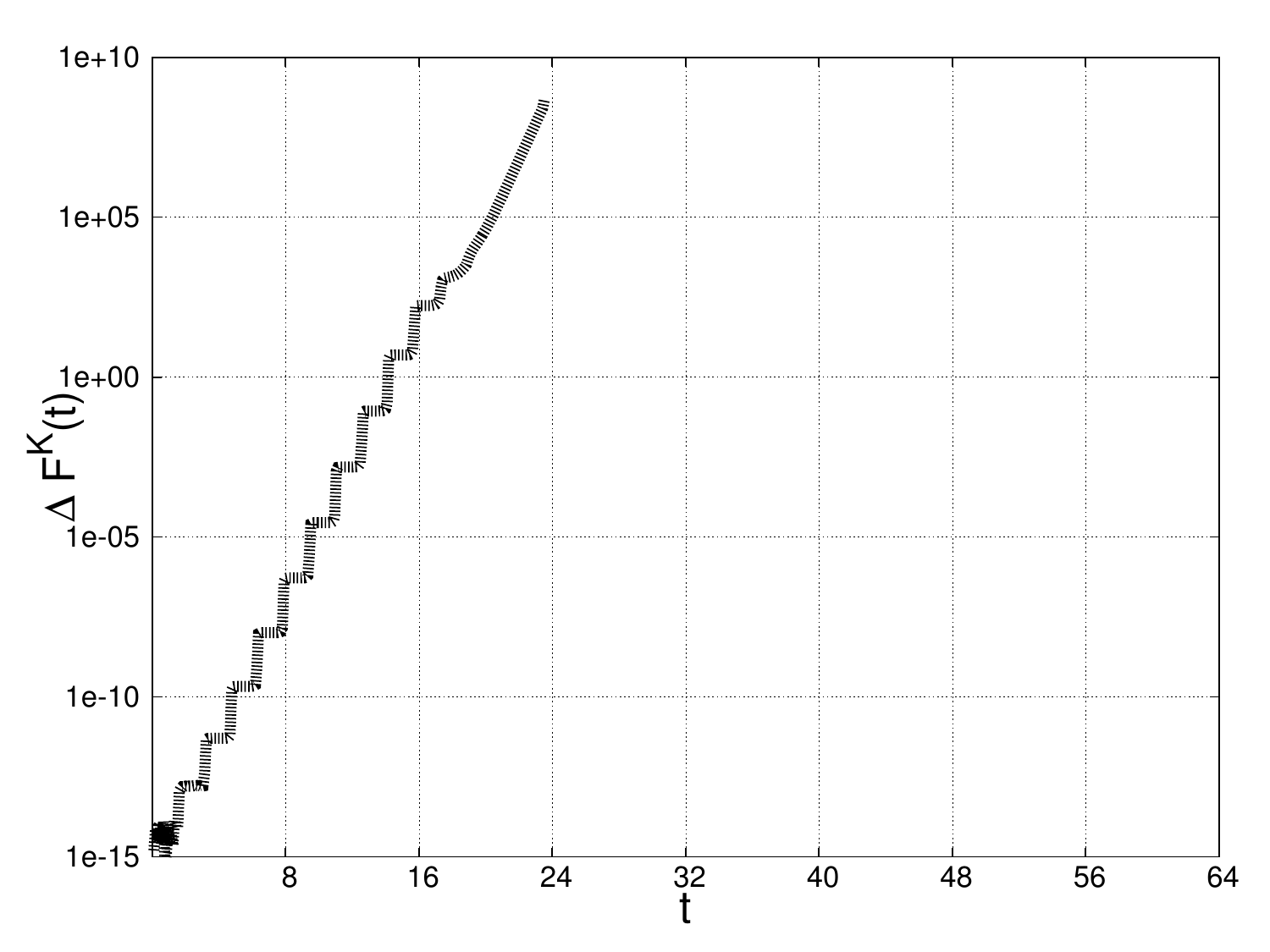}
  \caption{The relative error $\Delta F^K$ of the free energy as a function of integration time $t$ for the explicit Cash--Karp scheme in semilogarithmic scale.
  The left panel shows the result for $L=32$, 
  the right panel depicts the corresponding result for $L=64$.}
  \label{fig-deltaE-CK}
 \end{figure}

 In order to show, as a last point, why standard integration schemes are of no use for the discrete BCS equations, we apply the popular fifth order Cash--Karp scheme
 of~\cite{1992F&PT..8..NR}
 to the equations with the same $L$ and the same step size as for the splitting methods. When plotting the resulting $\Delta F^K$, cf.~Fig~\ref{fig-deltaE-CK},
 we observe an exponential growth in the error. This is in accordance with theoretical expectations, see, e.g.~\cite{1993H&NW..8..679M}.
 
 Let us now summarize our results.

\section{Conclusion}\label{IntSec:Con}
In this work, we have presented two fast and accurate evolution schemes, BCSInt and SplitBCS, for the coupled 
discrete BCS equations which arise from a Fourier space discretization of the BCS equations for 
superconducting materials. 
BCSInt uses the preservation of the density matrix' eigenvalues to decouple the system and a subsequent splitting
of the decoupled system into two terms.
SplitBCS is based on a splitting of the coupled equations into three subproblems which for the important
case of a contact interaction
can all be solved exactly by employing basic operations only. Crucially, the CPU effort for these
exact solutions grows only linearly in the dimension of the spatial discretization. 
Further computational costs could be saved by aptly recombining the flows of the subproblems. 
In numerical tests, the schemes have been shown to be very accurate. 
Additionally, they preserve the discrete analog of the physical energy and the eigenvalues of the particle density matrix
up to very small 
errors. We have, thus, come up with very useful tools for simulations in the field of superconductivity. 
%%%%%%%%%%%%%%%%%%%%%%%%%%%%%%%%%%%%%%%%%%%%%%%%%%%%%%%%%%%%%%%%%%%%%%%%%%%%%%%%%%%%%%
\begin{acknowledgements} I would like to thank Ch. Hainzl and Ch. Lubich for useful discussions and suggestions. This work was partially funded by the DFG grant GRK 1838.
\end{acknowledgements}

%%%%%%%%%%%%%%%%%%%%%%%%%%%%%%%%%%%%%%%%%%%%%%%%%%%%%%%%%%%%%%%%%%%%%%%%%%%%%%%%%%%%%%
 
\end{document}